\documentclass[a4paper]{jpconf}
\usepackage{graphicx}
\usepackage{enumitem}
\usepackage{tabularx}
\usepackage{makecell}
\usepackage{comment}

\begin{document}
\title{Evaluating Portable Parallelization Strategies for Heterogeneous Architectures in High Energy Physics}

\author{
Mohammad Atif$^1$, 
Meghna Battacharya$^2$,
Paolo Calafiura$^3$, 
Taylor Childers$^4$, 
Mark Dewing$^2$, 
Zhihua Dong$^1$, 
Oliver Gutsche$^2$, 
Salman Habib$^4$, 
Kyle Knoepfel$^2$, 
Matti Kortelainen$^2$, 
Ka Hei Martin Kwok$^2$, 
Charles Leggett$^3$, 
Meifeng Lin$^1$, 
Vincent Pascuzzi$^1$, 
Alexei Strelchenko$^2$, 
Vakhtang Tsulaia$^3$, 
Brett Viren$^1$, 
Tianle Wang$^1$, 
Beomki Yeo$^3$, 
Haiwang Yu$^1$
}

\address{$^1$ Brookhaven National Laboratory, Upton, NY 11973, USA }
\address{$^2$ Fermi National Accelerator Laboratory, Batavia, IL 60510, USA }
\address{$^3$ Lawrence Berkeley National Laboratory, Berkeley, CA 94720, USA }
\address{$^4$ Argonne National Laboratory, Lemont, IL 60439, USA }
\newcommand{\ptor}{\texttt{p2r}}


\begin{abstract}
High-energy physics (HEP) experiments have developed millions of lines of code over decades that are optimized to run on traditional x86 CPU systems. However, we are seeing a rapidly increasing fraction of floating point computing power in leadership-class computing facilities and traditional data centers coming from new accelerator architectures, such as GPUs. HEP experiments are now faced with the untenable prospect of rewriting millions of lines of x86 CPU code, for the increasingly dominant architectures found in these computational accelerators. This task is made more challenging by the architecture-specific languages and APIs promoted by manufacturers such as NVIDIA, Intel and AMD. Producing multiple, architecture-specific implementations is not a viable scenario, given the available person power and code maintenance issues.

The Portable Parallelization Strategies team of the HEP Center for Computational Excellence 
is investigating the use of Kokkos, SYCL, OpenMP, std::execution::parallel and alpaka as potential portability solutions that promise to execute on multiple architectures from the same source code, using representative use cases from major HEP experiments, including the DUNE experiment of the Long Baseline Neutrino Facility, and the ATLAS and CMS experiments of the Large Hadron Collider. This cross-cutting evaluation of portability solutions using real applications will help inform and guide the HEP community when choosing their software and hardware suites for the next generation of experimental frameworks. We present the outcomes of our studies, including performance metrics, porting challenges, API evaluations, and build system integration.
\end{abstract}

\section{Introduction}

High Energy Physics is facing an enormous challenge in the coming decades as data volumes and processing requirements for precision physics analysis and simulation increase dramatically with experiments such as the Deep Underground Neutrino Experiment (DUNE) and those on the High-Luminosity Large Hadron Collider (HL-LHC). Traditionally, computing for HEP has been undertaken at a mixture of institutional clusters, distributed grid sites, and HPC centers. Recently there has been increased use of commercial cloud resources, but this fraction is still small. All these computing resources have been almost entirely comprised of the same hardware architecture: x86-based CPUs. However, as HPC centers attempt to increase their computational power, energy consumption limits of this architecture have necessitated the introduction of GPU-based accelerators to provide the majority of the computing. NVIDIA, AMD and Intel GPUs are now being heavily used in recent and next generation HPC centers. 

Programming techniques for GPUs and other massively parallel accelerators are very different from that of traditional CPUs, requiring a significant understanding of the hardware architecture to achieve good performance. Special languages and compilers have been developed to target these architectures, such as CUDA for NVIDIA GPUs, HIP for AMD GPUs, and SYCL for Intel GPUs. 

HEP experiments have developed millions of lines of code, and in order to use these new computational accelerator facilities would need to rewrite large amounts of their code bases. In order to be able to run on all the different hardware architectures, this task would have to be repeated multiple times in each architecture's preferred language. This exercise, and the validation necessary to keep all versions coherent is both time and cost prohibitive. A solution needs to be found where code can be written once, and then run on all available architectures without modification. In recent years, a number of portability layers have been developed which address this problem, and the
ANONYMOUS group was created in order to test and evaluate these portability layers in relation to their usage in current and future HEP experiments.

The work presented in this paper represents one of the largest-scale portability studies using multiple programming models and portability layers in a diverse set of HEP applications. In addition to the achieved computational performance across different compute architectures, we also evaluate the different portability approaches using a comprehensive set of metrics most relevant to HEP software, such as modifications needed to the existing code, event data model, build system, etc. In this paper, we describe the portability layers studied (Section~\ref{sec:layers}), the representative HEP testbeds (Section~\ref{sec:testbeds}), metrics considered (Section~\ref{sec:metrics}),  and performance evaluations of different portability layers (Section~\ref{sec:perf}). We end the paper with a summary of our non-performance evaluations of different programming models in Section~\ref{sec:eval}. 

\section{Portability Layers}\label{sec:layers}

\begin{figure}[h]
\begin{center}
\includegraphics[width=0.8\textwidth]{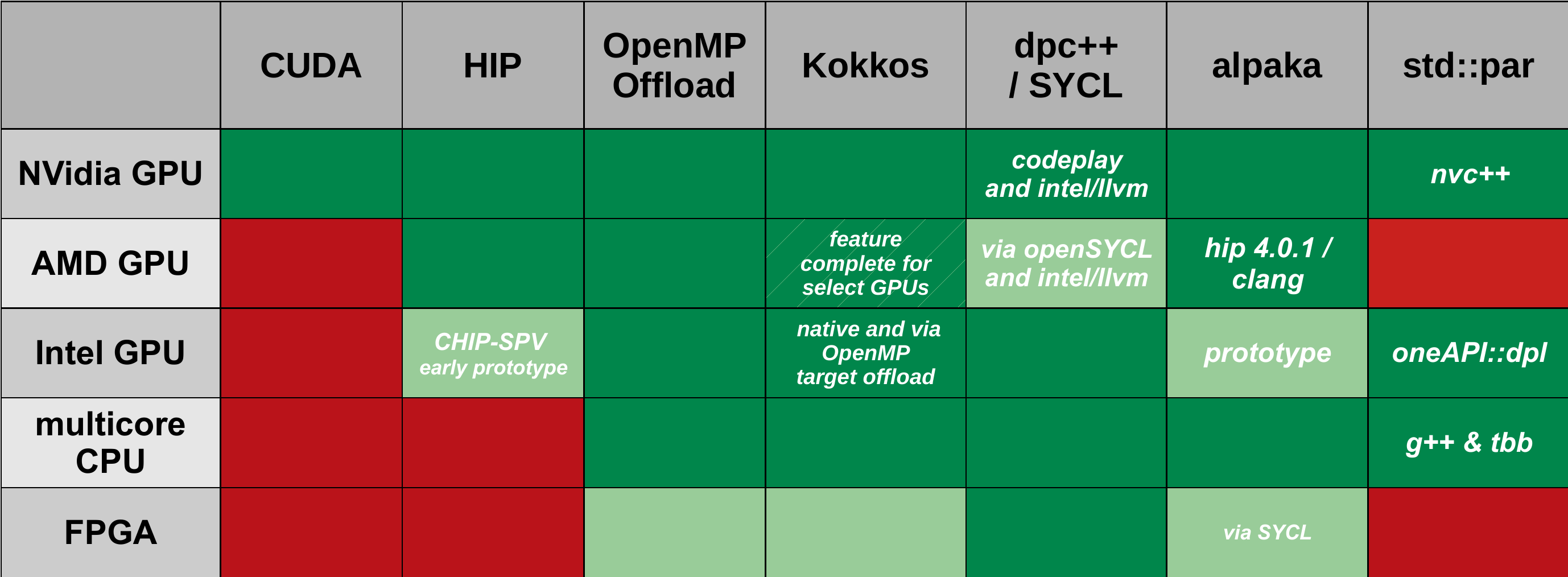}
\caption{\label{PL_matrix}Hardware support of portability layers. Dark green indicates full support, light green indicates partial support or that the project is still under development, and red indicates no support.}
\end{center}
\end{figure}

There are a number of currently existing portability layers (Figure~\ref{PL_matrix}) which permit user code to execute on various accelerator hardware architectures with little or no modifications to the source code. In general, selecting a new architecture will require a re-compilation of the code, though in certain cases an entirely different compiler must be used. These layers usually provide their own memory and data management facilities to attempt to abstract away the specifics of the backend hardware. We have chosen to evaluate Kokkos, SYCL, OpenMP, alpaka, and std::execution::parallel.
In this study, we have used CUDA as a baseline standard and comparison metric, as it is the most widely used GPU programming language in the field, and several of our testbeds already had pre-existing CUDA versions.

\subsection{Kokkos}
Kokkos~\cite{edwards:2014,trott:2022} is a portable, performant, C++ based shared-memory programming model that is single source, ie it lets you write algorithms once and run on any supported  backend architectures, such as a traditional CPU, NVIDIA, AMD, and Intel GPUs, and manycore CPUs, minimizing the amount of architecture-specific implementation details that users need to know. It provides a number of different parallel abstractions available, such as parallel\_for, reductions, and scans, and also provides utilities such as random number generators, and support for atomic operations, chained kernels and callbacks.
The library, which is mostly header based, is compiled for a selected set of backends - one serial, one host parallel, and one specific accelerator device can be chosen in a single binary. These backends must be selected at compile time. Though it provides constructs for allocating and managing data on the host and accelerator devices, these can be wrapped around pre-existing data objects. Execution kernels can also use bare pointers to operate on data allocated and transferred by other means. 

\subsection{SYCL}
SYCL is a cross platform abstraction layer intended for heterogeneous computing, based on OpenCL, and originally released by the Khronos group in 2014. Since then, there have been a number of implementations by different groups.
Like Kokkos, it is also single source, and understands C++17.
It does not mandate explicit memory transfers, but rather builds a DAG of kernel data dependencies, and transfers the data between host and offload device as needed. SYCL runs on a broad range of architectures, and in theory permits the selection of the execution devices at runtime. In practice, different accelerator backends require different compilers, such as openSYCL to target AMD GPUs, and different builds of llvm/dpc++ to target NVIDIA or Intel GPUs.

\subsection{OpenMP and OpenACC}

OpenMP is a very large and complex directive based API that enables multi-platform shared memory multiprocessing, with support for C, C++ and Fortran. It provides parallelism within a node and is heavily used for multicore and many-node CPUs in HPCs. In its recent versions, OpenMP has been extended to GPUs via the “target offloading” model which offloads a kernel to NVIDIA, AMD, and Intel devices. OpenACC is a similar standard that was developed explicitly for accelerator devices like GPUs. It allows the compiler to make intelligent decisions on how to decompose problems, and describes what the compiler \textit{should} do, not must. This allows different compilers to interpret "should" very differently.

\subsection{std::execution::parallel}

\verb|std::execution::parallel| (\verb|std::par|), is an existing C++ standard that was introduced in C++17 to enable parallel processing of algorithms by defining execution policies. Available policies are serial, parallel execution using threads, and parallel execution using threads and vectorization. In 2020 NVIDIA introduced a compiler (nvc++) which enabled the execution of parallel execution policies on NVIDIA GPUs. nvc++ uses unified shared memory for data access, with data being migrated to the GPU on demand via page faults. It is not intended to be a replacement for CUDA, as it lacks many low-level features and optimizations, but rather as a stepping stone or bridge between CPU-based serial code, and that explicitly written for GPUs, dramatically lowering the entry bar for execution on accelerators. Intel also has a compiler (oneAPI:dpl) that mostly supports this standard.

\subsection{alpaka}

alpaka~\cite{worpitz:2015,zenker:2016,mathes:2017} is another single source portability layer, where the backend is chosen at compile time.
It uses a CUDA-like multidimensional set of work units, using a grid / block / thread / element hierarchy, mapping the abstraction model to the desired backend.
It has a data-agnostic memory model, allocating  memory using smart memory buffers, which take care of proper memory handling for specific backends. Alpaka uses the same API to allocate memory on the host and on the device.
Alpaka also offers straightforward porting of CUDA kernels using its extension called "cupla", where only the includes and syntax of the kernel calls need to be changed.

\section{Testbeds} \label{sec:testbeds}

In order to evaluate the various portability layers, we have selected a number of representative testbeds from several HEP experiments, including ATLAS, CMS and DUNE. These testbeds are relatively small code bases that can be executed outside of the experiments' full frameworks in order to simplify the development process. Some of the testbeds had already been ported from the original serial code to run on GPUs using CUDA. We have used either the original serial CPU code or the CUDA versions to develop implementations for all of the different portability layers that we are exploring.

\subsection{Patatrack}
Patatrack is a standalone version~\cite{kortelainen:2021a} of CMS heterogeneous pixel reconstruction~\cite{bocci:2020b}. The chain of reconstruction algorithms takes the raw data of the CMS pixel detector as an input, along with the beamspot parameters and necessary calibration data, and produces pixel tracks and vertices. The algorithms are divided into about 40 kernels, and were originally implemented in CUDA. The standalone setup includes a multithreaded testing framework mimicking relevant aspects of the CMS data processing framework, CMSSW~\cite{jones:2006,jones:2014,jones:2015,jones:2017,bocci:2020a}. Many of the kernels and memory copies are short, and therefore the application is very sensitive to overheads. The application attempts to maximize the GPU utilization by processing events concurrently with CUDA streams, leveraging asynchronous execution capabilities of the CMSSW framework and minimizing synchronization points, and using a memory pool between the algorithms and CUDA runtime API.

\subsection{Wire-Cell Toolkit}
Wire-Cell Toolkit (WCT) is a new standalone C++ software package for Liquid Argon Time Projection Chamber (TPC) simulation, signal processing, reconstruction and visualization, and is intended to be used by the Deep Underground Neutrino Experiment (DUNE). It is written in modern C++, following the data flow programming paradigm. It supports both single-threaded and multi-threaded execution on CPUs with the choice determined by user configuration. 
WCT currently includes central elements for DUNE data analysis, such as signal and noise simulation, noise filtering and signal processing, and is currently deployed in production jobs for MicroBooNE and ProtoDUNE experiments. 

For the portability evaluation, we chose the Liquid Argon Time Projection Chamber (LArTPC) signal simulation module in WCT, as it is computationally expensive and may benefit from acceleration on GPUs. The LArTPC signal simulation is composed of three major steps: ``rasterization'' to decompose data (individual ionization electron groups) into patches of various sizes, ``scatter-add'' to sum up the patches into a larger grid, and ``convolution'' to obtain the simulated detector signal. The rasterization step involves nested for loops that can be parallelized and offloaded to GPUs. Scatter-add requires \texttt{atomic} operations, while convolution's main computation is done through Fast Fourier Transforms (FFT). The three steps make up the majority of the \texttt{DepoTransform} routine, which is the main computational kernel for the LArTPC signal simulation. In serial execution on CPUs, rasterization is typically the most time-consuming step, followed by convolution. 

\subsection{FastCaloSim}
The ATLAS detector at the Large Hadron Collider (LHC) relies on large samples of simulated proton-proton collision events for detector modeling, upgrade studies and to deliver high-quality physics results.
Standard simulations using Geant4~\cite{Agostinelli:2002hh} for accurate and detailed modeling of physics processes and detector geometry are extremely CPU-intensive, and its use for all Monte Carlo-based studies is impractical due to the formidable size of the LHC dataset.
In particular, simulating electromagnetic and hadronic shower developments in the ATLAS calorimeters is extremely CPU-intensive, comprising more than 90\% of the overall simulation time~\cite{ATLAS_Sim_Infrastructure}.

To reduce the significant processing time required to simulate the calorimeters,  ATLAS developed FastCaloSim~\cite{ATL-SOFT-PUB-2018-002} which uses a simplified detector geometry and parameterizations of shower development initiated by particles traversing the calorimeter volumes.
During a simulation, the optimal parameterization is selected based on the particle type, energy deposit and location in the detector in order to best model the corresponding electromagnetic or hadronic shower.
This parameterized simulation can reduce the CPU time by a factor of 10 to 25 — depending on the process being modeled — relative to that of fully Geant4-based simulations. 

The main event loop of the FastCaloSim testbed is comprised of four parts: 1) workspace initialization, where a large array representing the calorimeter cells on the GPU is reset, 2) the main simulation where energy deposits of the particle hits are calculated, 3) stream compaction where the hits are combined to determine the energy deposits in the calorimeter, and 4) copying the results from the GPU to the host.
There are two different versions of the code: the first simulates the detector response a particle at a time, and the second groups multiple particles together before offloading the calculations to the GPU. Offloading a single particle at a time makes inefficient use of the GPU, as the calculations may not be numerically intensive, and can often be limited by the launch latency of the GPU. Grouping particles together results in much larger workloads for the GPU, improving GPU efficiency. 
Details can be found at \cite{ATLAS_FCS_GPU}.

\subsection{p2r}
Propagation-to-r~(\ptor)~\cite{ptor} is a standalone mini-application, which performs the core math of parallelized track reconstructions. The kernel aims at building charged particle tracks in the radial direction under a magnetic field from detector hits, which involves propagating the track states and performing Kalman updates after the propagation.  The kernels are implemented based on a more realistic application, called $\texttt{mkFit}$~\cite{mkfit}, which performs vectorized CPU track fitting. 
To simplify the track reconstruction problem, the $\ptor$ program processes a fixed number of events with the same number of tracks in each events. All track computations are implemented in a single GPU kernel. A fixed set of input track parameters is smeared randomly and then used for every tracks.    

\section{Metrics} \label{sec:metrics}

Choosing between the different portability layers is not an easy task. Merely looking at their computational performance, while important, is not sufficient, as there are many other characteristics which may affect their adoption by an experiment. In order to evaluate them, we have identified a number of metrics that will provide both subjective and objective measures of their different characteristics and suitability for their end users.

Below are the simplified list of the 14 metrics that will be evaluated using each of the test-beds. The full metrics can be found in Ref~\cite{metric}. The first 5 metrics focus on the porting and development experience, whereas the rest of the metrics explore other advantages/disadvantages or potential limitations associated with the portability layer. 
Depending on the scale of the test-bed and nature of the computation, the evaluation on each metric could vary. With a representative set of test-beds, the evaluation aims to provide valuable information to different use-cases of the HEP communities. 

\begin{enumerate}
    \item Ease of Learning
    \item Code conversion
    \begin{itemize}
        \item From CPU to GPU and between different APIs
    \end{itemize}
\item Extent of modifications to existing code 
\begin{itemize}
    \item Control of main, threading/execution model
\end{itemize}
\item Extent of modifications to the Data Model
\item Extent of modifications to the build system
\item Hardware Mapping
\begin{itemize}
    \item Current and promised future support of hardware
\end{itemize}

\item Feature Availability
\item Address needs of large and small workflows
\item Long term sustainability and code stability
\begin{itemize}
    \item Backward/forward compatibility of API and e.g. CUDA
\end{itemize}
\item Compilation time
\item Run time/Performance
\item Ease of Debugging
\item Aesthetics
\item Interoperability
\begin{itemize}
    \item Interaction with external libraries, thread pools, C++ standards
\end{itemize}
\end{enumerate}

\section{Performance Studies} \label{sec:perf}
\subsection{Patatrack}

The standalone Patatrack code was first ported from CUDA to Kokkos in \cite{kortelainen:2021a}, and has since then been ported to HIP, and to Alpaka~\cite{bocci:2023}, and SYCL~\cite{perego:2022} by other groups. 

When benchmarking the Kokkos performance, we compare the event processing throughput using both the host and device execution spaces. As shown in Table \ref{patatrack:cpu}, the Kokkos host serial execution space is about 1.6x slower than the original CPU version when only a single thread is used, and when all 40 threads on a 20 core Intel Xeon Gold 6148 CPU are used, the Kokkos thread execution space is about 20x slower than the multithreaded CPU version. The multithreaded CPU version uses inter-event parallelism, where each event is processed in its own thread, whereas the Kokkos version attempts to parallelize \textit{within} a single event. Until very recently, Kokkos was not compatible with external use of thread pools such as TBB, and while it can now co-exist, the performance is still very poor.

The CUDA version of the code is able to process data from multiple events concurrently each in its own thread, with each thread owning its own CUDA stream. It also uses a memory pool to amortize the cost of raw CUDA memory allocations. For the closest comparison with the Kokkos version using the CUDA device parallel backend, we disable the CUDA memory pool and only allow a single concurrent event, at which point we see in Table \ref{patatrack:cuda} that the CUDA version is 37\% faster than the Kokkos version. It proved infeasible to implement a comparable device memory pool in Kokkos. If we enable the memory pool, the performance gap increases to 6.2x. And if we allow the CUDA version to process multiple concurrent events via multithreading, this increases yet again such that the CUDA version with 9 concurrent events is 16x faster than the Kokkos. Increasing the number of concurrent events beyond 9 shows no improvement in throughput.

The std::par port of the Patatrack is complete, however bugs in the nvc++ compiler prevent it from running. Similarly, OpenMP offload is still lacking some features that are used in the CUDA version, making the OpenMP port incomplete. We await the continued improvement of both these compilers in order to complete our benchmarks.

\begin{table}[b]
\centering
\caption{\footnotesize Comparison of the event processing throughput between the Kokkos version of Patatrack using Serial and Threads execution spaces and the CPU version implemented from the original CUDA version through a simple translation header. In all cases all the threads were pinned to a single CPU socket (Intel Xeon Gold 6148) that has 20 cores and 2 threads per core. Each test ran about 5 minutes, and CPU-heavy threads from a background process were used to fill all the 40 hardware threads of the socket. The work in the CPU version is parallelized by processing as many events concurrently as the number of threads the job uses without any intra-event parallelization, whereas in the Kokkos version there is only one event in flight, and all parallelization is within the data of that event. For the Kokkos version with Threads execution space the maximum throughput from a scan from 1 to 20 threads is reported. The reported uncertainty corresponds to sample standard deviation of 8 trials. From Ref.~\cite{kortelainen:2021a}}
\label{patatrack:cpu}       
\begin{tabularx}{0.48\textwidth}{
>{\raggedright\arraybackslash\hsize=5cm}X
>{\centering\arraybackslash}X
}
\hline
Test case & Throughput (events/s) \\
\hline
CPU version, 1 thread & $13.5 \pm 0.2$ \\
Kokkos version, Serial execution space & $8.5 \pm 0.2$ \\
\hline
CPU version, 40 threads & $539\pm 9$ \\
Kokkos version, Thread 
execution space, peak (18 threads) & $28 \pm 1$ \\
\hline
\end{tabularx}
\end{table}

\begin{table}[tb]
\centering
\caption{\footnotesize Comparison of the event processing throughput between the Kokkos version of Patatrack using CUDA execution space and the original CUDA version. In all cases the CPU threads were pinned to a single CPU socket, and used one NVIDIA V100 GPU. Each test ran about 5 minutes, and the machine was free from other activity. The CUDA version can process data from multiple events concurrently using many CPU threads and CUDA streams, and uses a memory pool to amortize the cost of raw CUDA memory allocations. The maximum throughput from a scan from 1 to 20 concurrent events is reported for the CUDA version. In order to compare to the current state of the Kokkos version, the CUDA version was tested also with 1 concurrent event and disabling the use of the memory pool. The reported uncertainty corresponds to sample standard deviation of 8 trials. From Ref.~\cite{kortelainen:2021a}}
\label{patatrack:cuda}       
\begin{tabularx}{0.48\textwidth}{
>{\raggedright\arraybackslash\hsize=5cm}X
>{\centering\arraybackslash}X
}
\hline
Test case & Throughput (events/s) \\
\hline
CUDA version, peak (9 concurrent events and CPU threads) & $1840\pm 20$ \\
CUDA version, 1 concurrent event & $720 \pm 20$ \\
CUDA version, 1 concurrent event, memory pool disabled & $159 \pm 1$ \\
\hline
Kokkos version, CUDA execution space & $115.7 \pm 0.3$ \\
\hline
\end{tabularx}
\end{table}

\subsection{Wire-Cell Toolkit}
\begin{figure}[h]
\centering
\includegraphics[width=.8\textwidth]{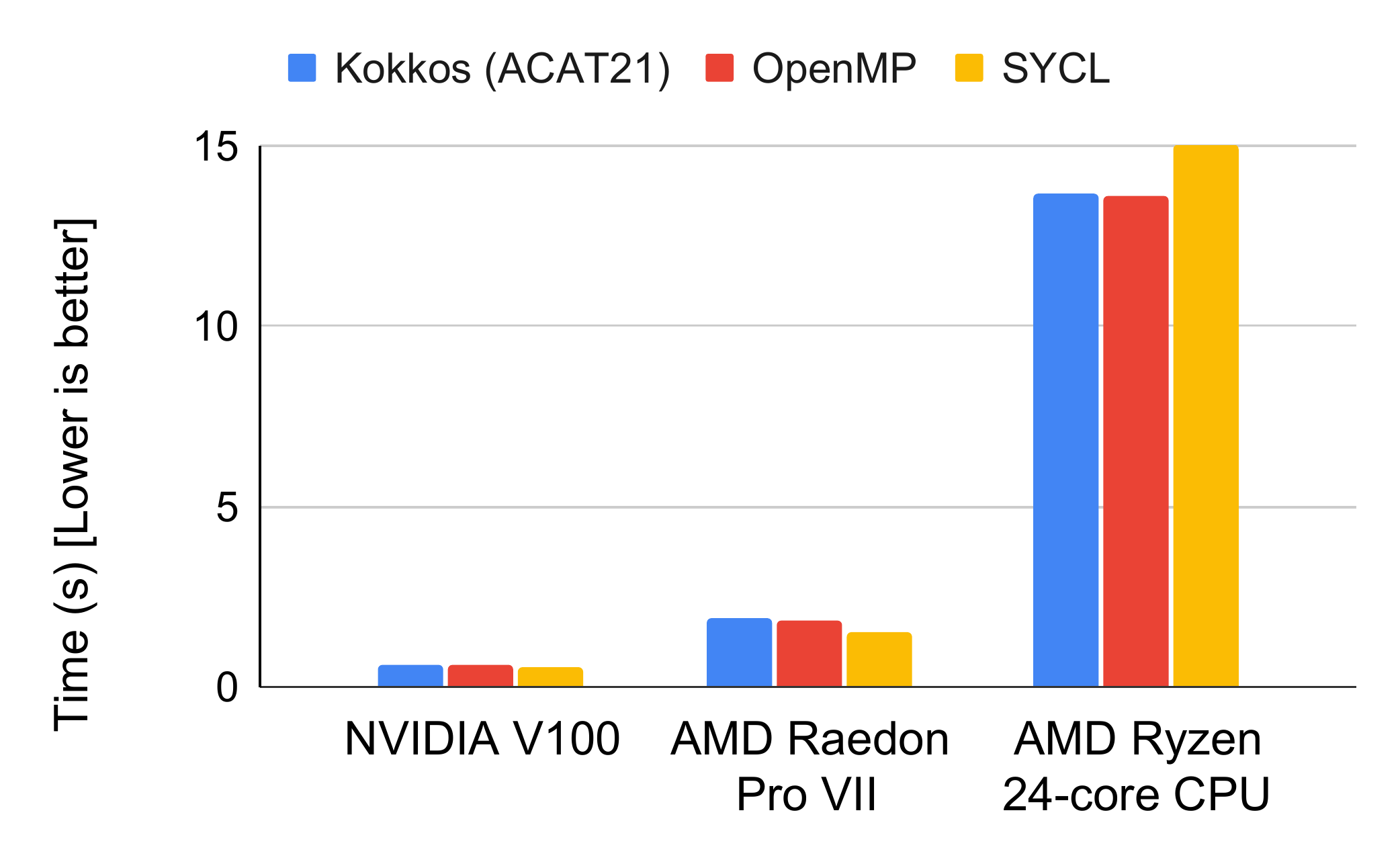}
\caption{\label{fig:wct}Performance comparison of Wire-Cell Toolkit \texttt{DepoTransform} kernel using Kokkos, SYCL and OpenMP on three different architectures. Timing was averaged over 20 runs in each case.}
\end{figure}

The first full GPU implementation for the Wire-Cell Toolkit (WCT) test case, signal simulation of the LArTPC detector,  was done with Kokkos ~\cite{yu2021evaluation, dong2023evaluation}, 
so we do not have a CUDA version as a baseline comparison. Since then, we have also ported the WCT test case to OpenMP and SYCL. Figure~\ref{fig:wct} compares the timing of the computational kernel \texttt{DepoTransform} in WCT using Kokkos, SYCL and OpenMP on three different architectures: NVIDIA V100 GPU, AMD Raedon Pro VII GPU and AMD Ryzen 24-core CPU. The compilers used for these platforms for each programming model are tabulated in Table~\ref{tab:wct-compilers}. While there are some small variations in the timing, all three programming models achieve very similar performance after some tuning and performance optimization. 

\begin{table}[htbp]
    \centering

    \begin{tabular}{c|c|c|c}
    \hline
    Target & Model & Compiler & Version   \\ 
    \hline
   NVIDIA GPU & Kokkos/3.3.01 & nvcc & 11.0.2  \\ 
              & {SYCL} & Intel/LLVM  & nightly-20220425  \\ 
              & OpenMP & LLVM/Clang & 15 \\ 
    \hline 
    AMD GPU & Kokkos/3.3.01 & rocm/hipcc & 4.5.2 \\ 
              & SYCL & Intel/LLVM  & nightly-20220425 \\ 
              & OpenMP & LLVM/Clang & 15 \\
    \hline
    AMD CPU & Kokkos/3.3.01 & GCC & 9.3.0 \\ 
            & SYCL & Intel OneAPI & 2022.02  \\ 
            & OpenMP & LLVM/Clang & 15 \\ 
    \hline 
    \end{tabular}
    \caption{ Compilers used for WCT portability studies. }
    \label{tab:wct-compilers}
    \end{table}

\subsection{FastCaloSim}

\begin{figure}[h]
\centering
\includegraphics[width=.8\textwidth]{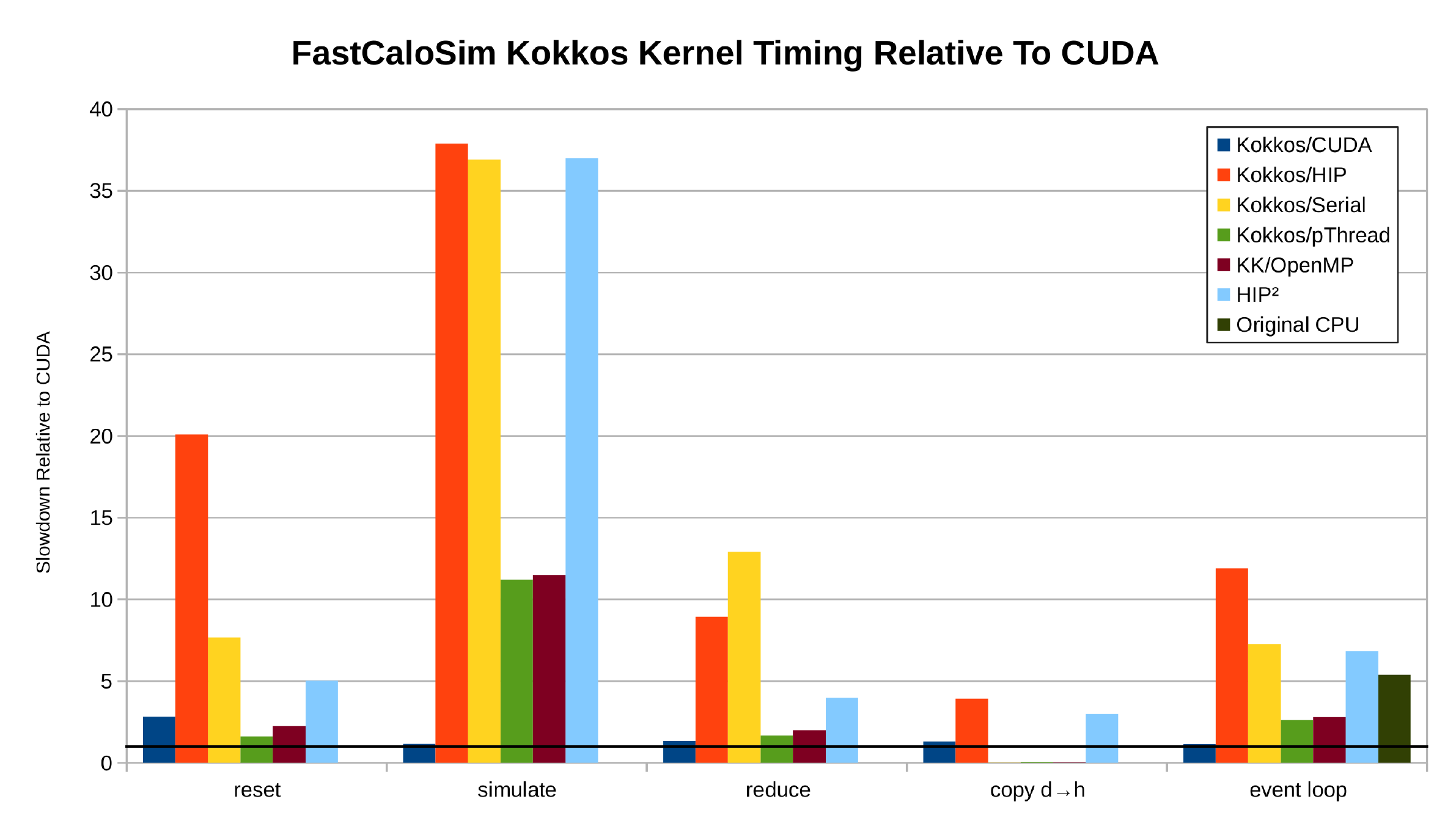}
\caption{\label{fig:fcs_kok}Performance of FastCaloSim using Kokkos relative to CUDA.}
\end{figure}

\begin{figure}[h]
\centering
\includegraphics[width=.8\textwidth]{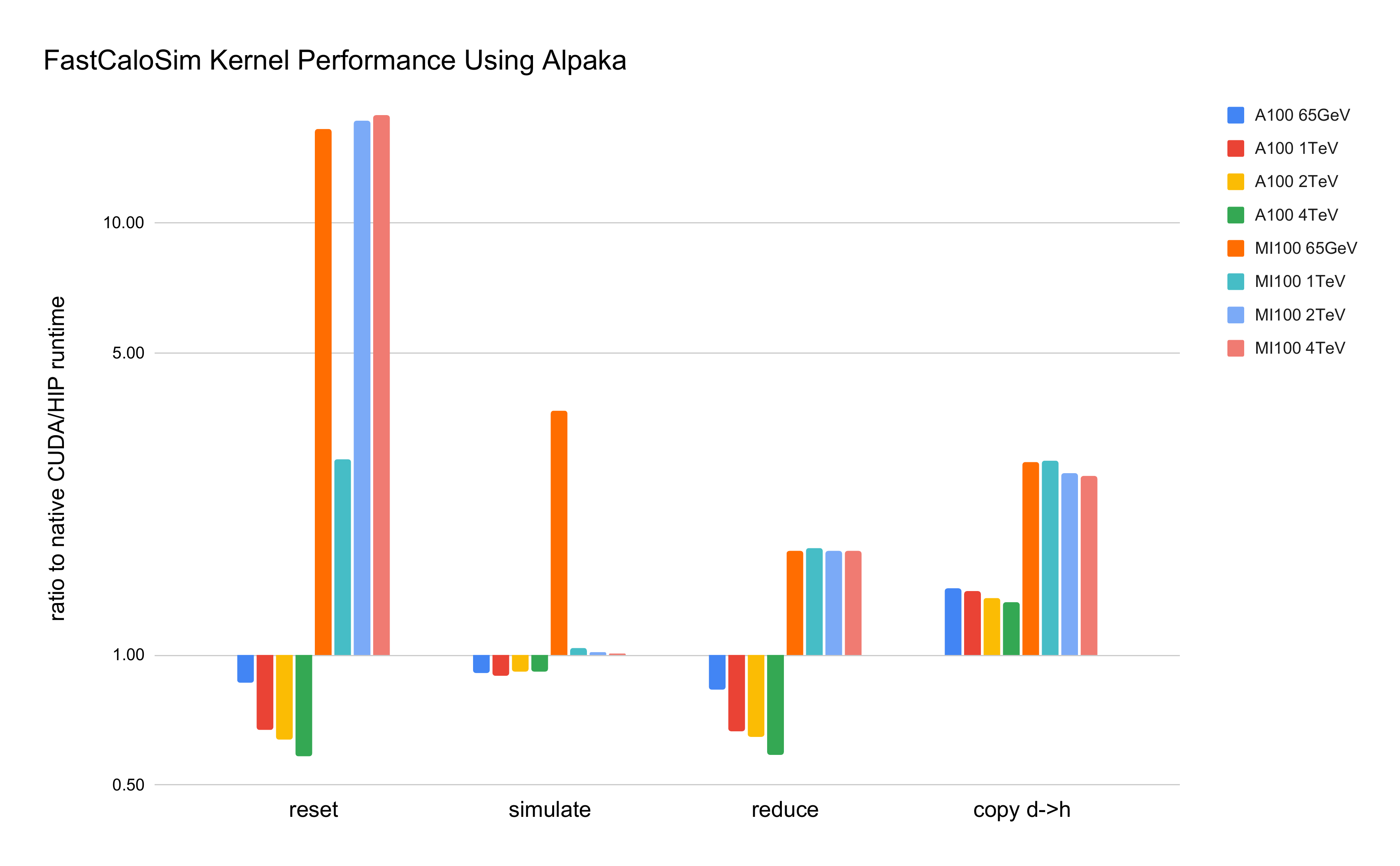}
\caption{\label{fig:fcs_alpaka}Performance of FastCaloSim using alpaka, relative to native CUDA/HIP.}
\end{figure}

\begin{figure}[h]
\centering
\includegraphics[width=.8\textwidth]{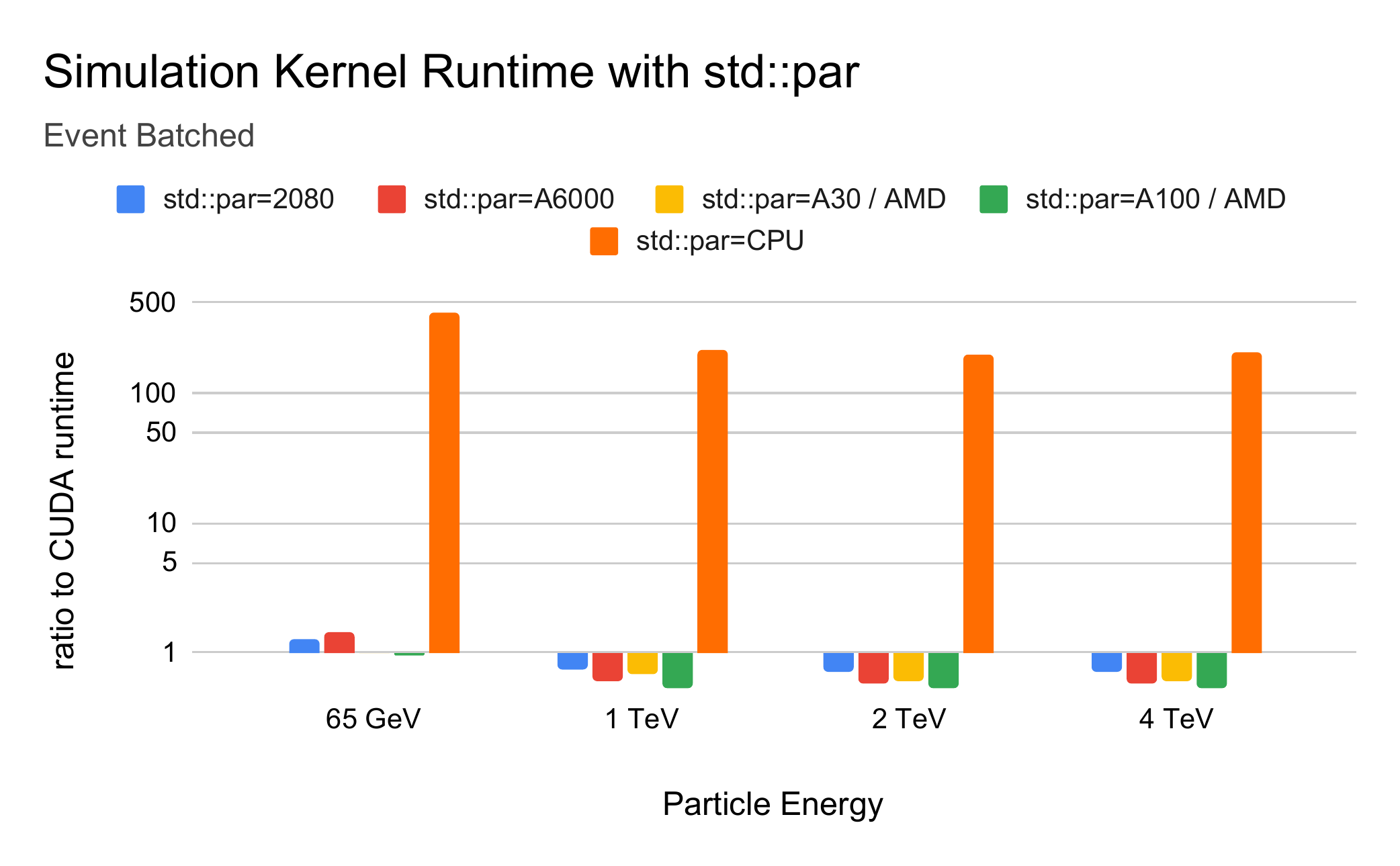}
\caption{\label{fig:fcs_stdpar_sim}Performance of FastCaloSim Simulation kernel for event batched data using std::par.}
\end{figure}

The original FastCaloSim serial CPU code was first ported to CUDA, and we have ported that version to Kokkos, SYCL, std::par, OpenMP and alpaka. 
We have used the CUDA implementation as a baseline to compare performance. A certain amount of code restructuring was necessary for the port to Kokkos, due to the lack of support for jagged arrays in Kokkos. These jagged data structures are flattened into 1D arrays or padded into regular 2D arrays. In general the portability backends tended to perform similarly to the original native CUDA or HIP implementation for the main simulation kernel execution as seen in Figures \ref{fig:fcs_kok}, \ref{fig:fcs_alpaka} and \ref{fig:fcs_stdpar_sim}. Overheads are often seen for the initialization, manipulation and transfers of the data when portability layers are used. 

The SYCL port has a substantially different code structure as compared with the CUDA and Kokkos ports, making comparisons of the individual kernel elements more challenging. In general however, the main simulation throughput is similar to that of CUDA, and SYCL does suffer from as many initialization overheads as Kokkos.

The alpaka port is structurally very similar to Kokkos, and while the performance of the main kernel is comparable to the native CUDA or HIP implementations, we do see some unusual behavior for the workspace resetting, which is still not fully understood. It seems to be related to how alpaka does device synchronization. We also see some small overheads from memory transfers.

\begin{figure}[h]
\centering
\includegraphics[width=.8\textwidth]{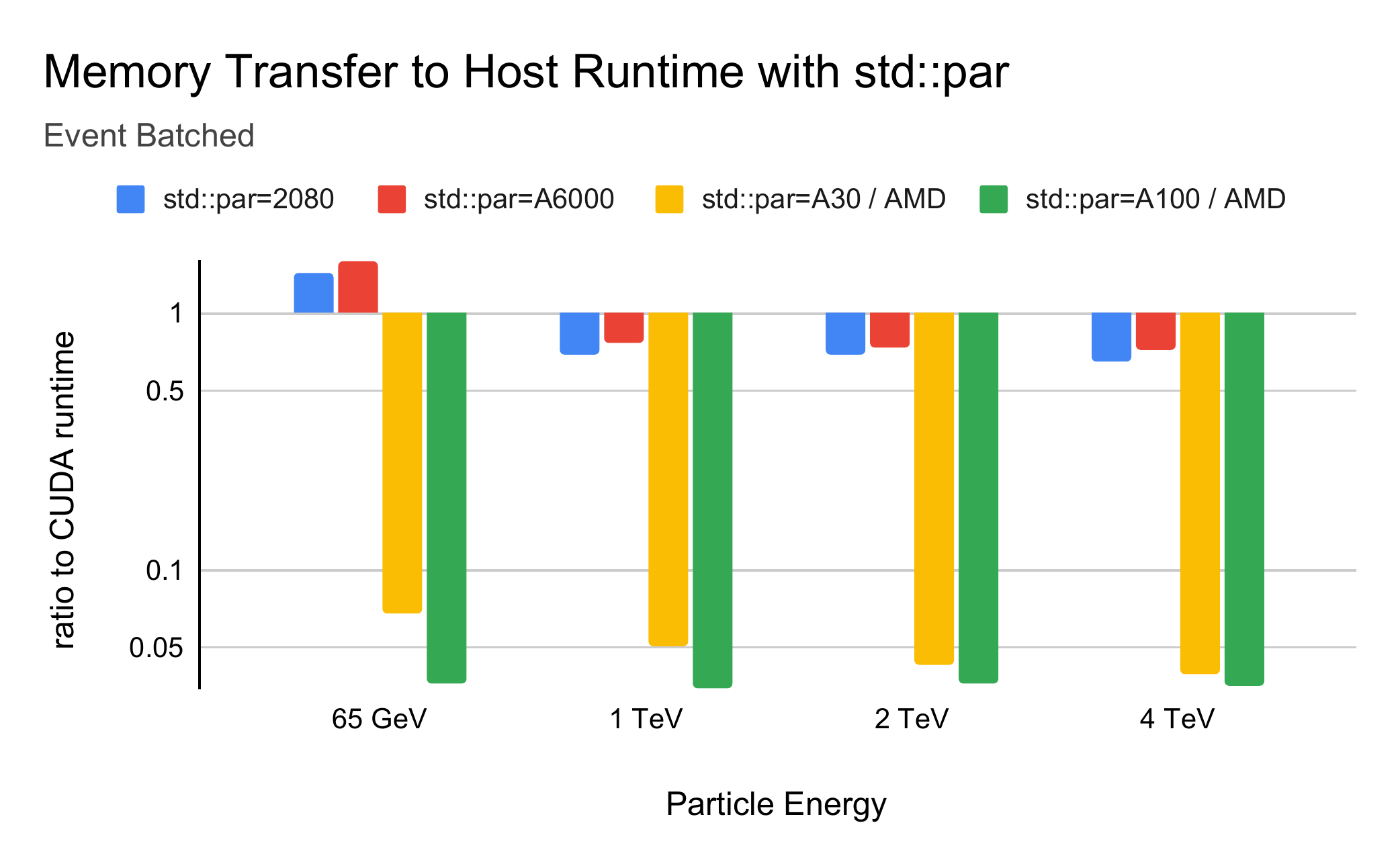}
\caption{\label{fig:fcs_stdpar_cpd2h}Performance of FastCaloSim data transfers with event batching using std::par.}
\end{figure}

One very unusual feature that was noted is that data transfers from the device to the host when using std::par are significantly degraded when the host CPU is an AMD EPYC (see Figure \ref{fig:fcs_stdpar_cpd2h}). The cause is thought to be related to how page faults are handled by the hardware, but the exact nature is still under investigation by NVIDIA. Another interesting observation is that the Thrust implementation of std::par is, in certain circumstances, faster than the original CUDA code. As well, when using the serial single core CPU backend of std::par, the overall performance is 20\% faster than the original serial CPU code. We have not been able to exercise the multicore backend of nvc++ due to compiler bugs.

\begin{figure}[h]
\centering
\includegraphics[trim={0cm 8.cm 18cm 0cm},clip,width=.8\textwidth]{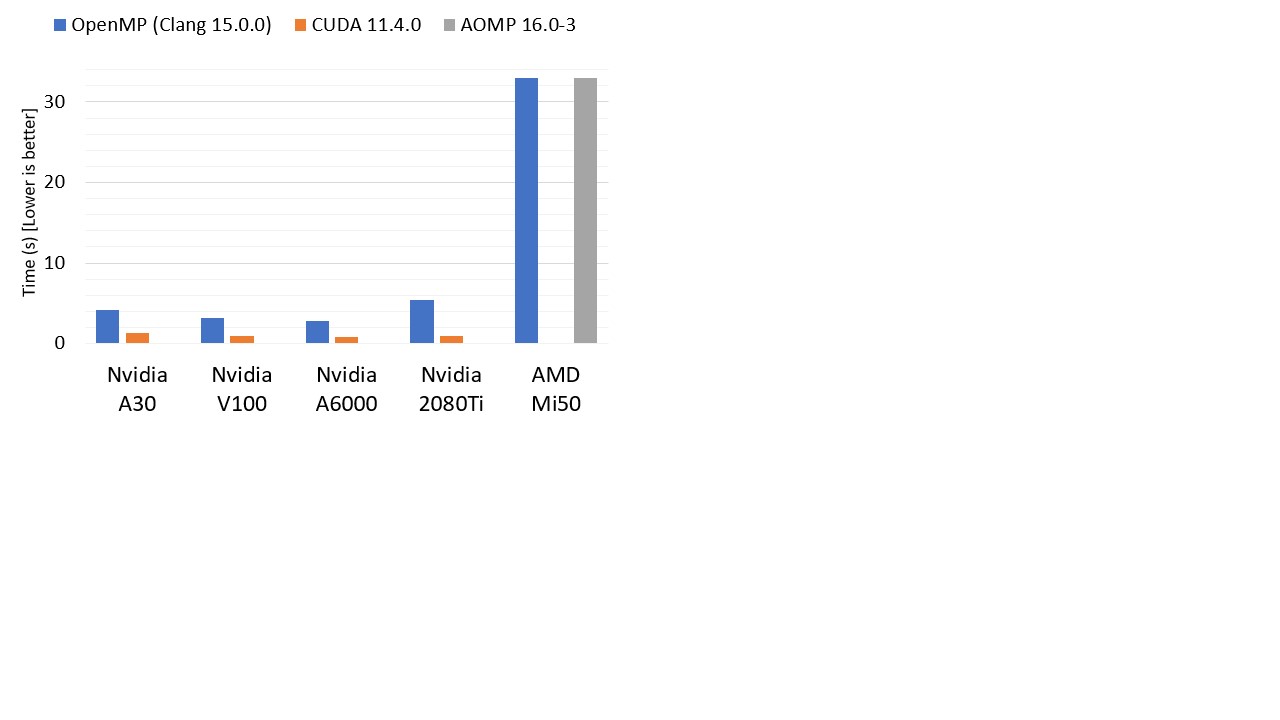}
\caption{\label{fig:fcs_omp} Performance of FastCaloSim Simulation kernel using OpenMP compared with CUDA.}
\end{figure}

\subsection{p2r}
$\ptor$ has been ported to alpaka, Kokkos, SYCL and std::par, and we have evaluated the performance of the GPU backends on a NVIDIA A100 and an AMD MI100 GPU, using the Joint Laboratory for System Evaluation (JLSE) system at Argonne National Laboratory. 

\begin{figure}[h]
\centering
\includegraphics[width=0.8\textwidth]{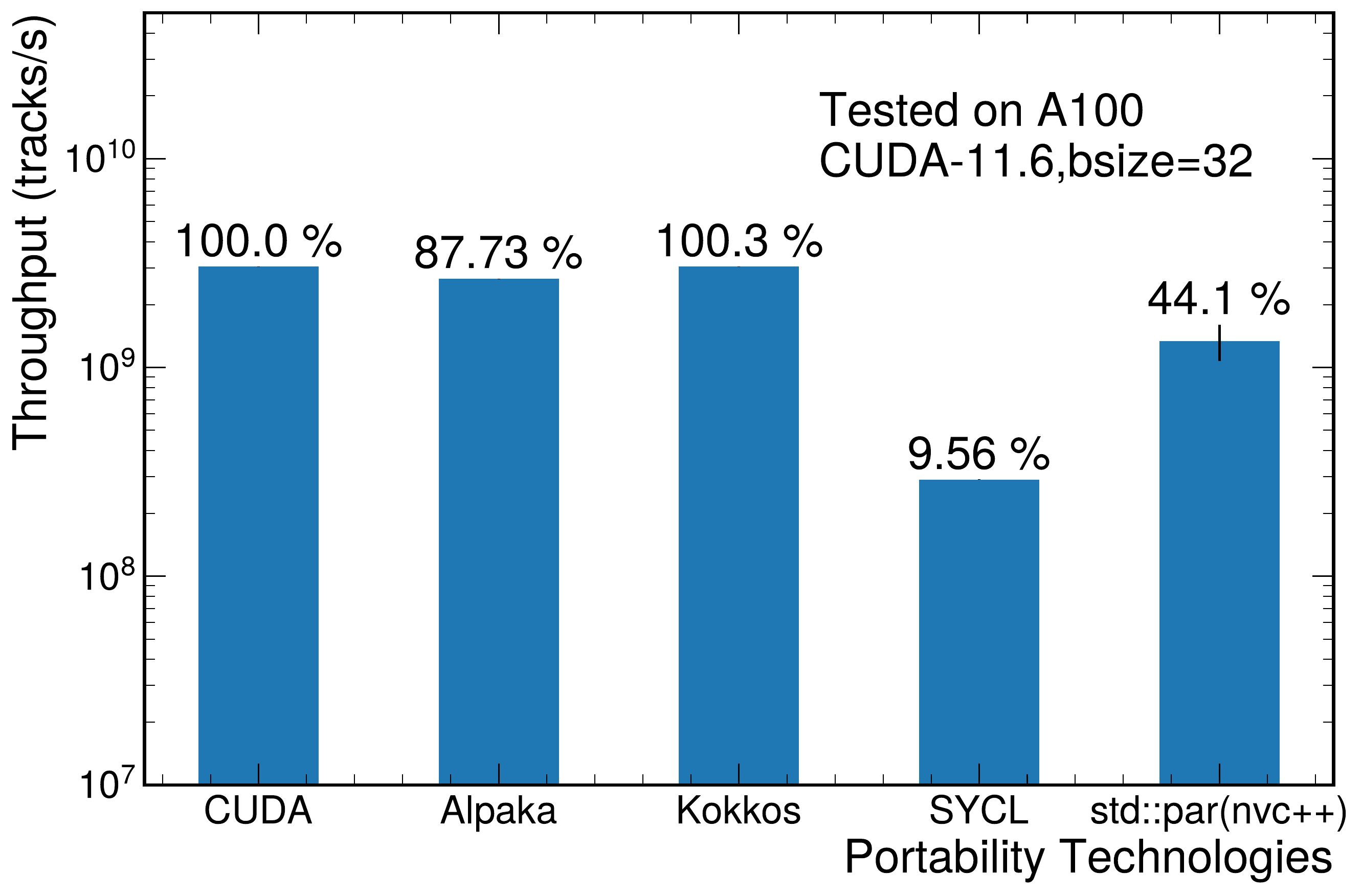}
\caption{\label{fig:p2r-cuda} Performance comparison of $\ptor$ using CUDA, alpaka, Kokkos, SYCL and std::par in NVIDIA A-100 GPU.}    
\end{figure}

\begin{figure}[h]
\centering
\includegraphics[width=0.8\textwidth]{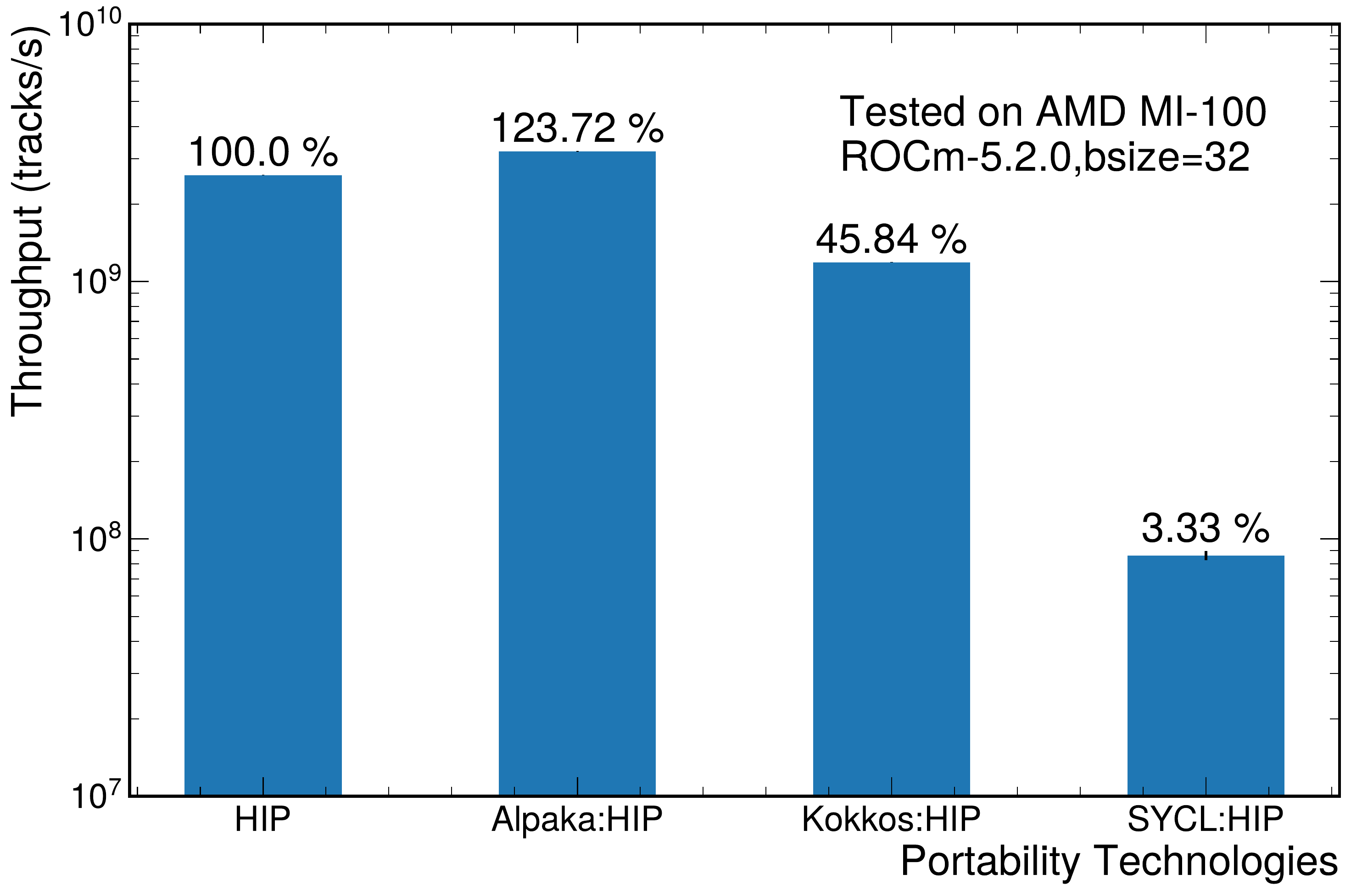}    
\caption{\label{fig:p2r-amd} Performance comparison of $\ptor$ using HIP, alpaka, Kokkos, SYCL in AMD MI-100 GPU.} 
\end{figure}

The kernel-only throughput achieved by using a portability layer is compared to that of the native programming model. On the A100 GPU, alpaka and Kokkos's performance is very close to the native CUDA version, whereas SYCL and std::par are approximately factor of 10 and 2 slower respectively.  We note, however, that both alpaka and Kokkos suffer from a $\sim40$\% slow down if the kernel launch parameters determined by the portability layer are used. Results shown in Figure~\ref{fig:p2r-cuda} and~\ref{fig:p2r-amd} are obtained by explicitly choosing the same launch parameters and register per thread values as the native CUDA version.

On an AMD MI100, alpaka achieved 23\% better performance than the native HIP implementation. While it is possible that the portability layer could introduce better optimization than the native implementation, further profiling work is required to confirm the cause for the better performance observed in this case. The poor performance of the SYCL implementation on both tested GPUs is not fully understood. Initial profiling results show orders of magnitude more instructions were executed in the kernel, inducing much more memory traffic and hence latency.

\section{Portability Layer Evaluations}
\label{sec:eval}

Over the past three years, all four testbeds have been ported to each portability layer, except for the alpaka port of the Wirecell Toolkit, for which there was insufficient person power. The std::par port of Patatrack, while complete, has encountered compiler bugs, for which we are still awaiting resolution from NVIDIA. A brief overview of the evaluation metrics is shown in Tables \ref{tab:metrics_1} and \ref{tab:metrics_2}, with some further details listed below.

\begin{table*}[ht]
\centering

\begin{tabularx}{0.99\textwidth}{ 
| >{\raggedleft\arraybackslash\hsize=2cm}X
| >{\raggedright\arraybackslash}X
| >{\raggedright\arraybackslash}X
| >{\raggedright\arraybackslash}X
| >{\raggedright\arraybackslash}X
| >{\raggedright\arraybackslash}X
|
}
\hline
\textbf{Metric} & 
\textbf{Kokkos} & 
\textbf{alpaka} & 
\textbf{SYCL} & 
\footnotesize \textbf{std::par/nvc\tiny++} & 
\textbf{OpenMP} \\
\hline

\small Ease of Learning& 
\tiny Similar to C++ and CUDA, optimization more challenging &
\tiny Very verbose API and sparse documentation make for steep learning curve &
\tiny Similar to C++ and CUDA. Lots of documentation. & 
\tiny is C++ &
\tiny C++ with extra pragmas. Sparse documentation/examples for offload. \\

\small Serial Code Conversion &
\tiny Similar to CUDA, though different syntax. 
Specialized optimizations not straightforward &
\tiny Work needed to wrap kernels in callable objects. Many typedefs in API benefit from layer of template functions. 
&
\tiny Similar to CUDA, though different syntax &
\tiny very simple &
\tiny Easy incremental porting mechanism from serial code (add pragmas, get offload working, performance tuning).   More work to port from existing CUDA code. 
\\

\small Code Modification &
\tiny Can be used in an existing complicated application without changes elsewhere.
Kokkos runtime needs explicit init/fini. 
Can take and interpret command line arguments &
\tiny similar to CUDA &
\tiny Parallelize loops with command group handlers and parallel\_for.
If using USM, necessary to orchestrate kernel calls explicitly via waiting on events &
\tiny Memory accessible on device must be allocated/freed in a file compiled by nvc++, and on the heap. This may require some copying of data. &
\tiny Special memory allocation and transfer APIs. Can operate device and host parallel simultaneously.
\\

\small Data Model Modification &
\tiny Views can be used as a smart pointer to 1D data
Crafting a SoA with Views tedious.
Jagged arrays (Views of Views) not gracefully supported &
\tiny Buffers used to wrap existing objects tedious to use. Alpaka managed memory buffers can lead to unexpected behavior &
\tiny Buffers can be instantiated only from device copyable types. 
USM is most compatible with current EDM and custom data types. &
\tiny May need to copy data to make it visible to USM &
\tiny In general simpler than CUDA
\\

\small Build System Integration &
\tiny Can choose at most one backend for each execution space type. Choice must be done at the time of configuring the Kokkos runtime build. &
\tiny Extensive configurability via CMake &
\tiny Mostly seamless integration with CMake and make. Depending on the target platform/backend, additional Clang command line arguments are needed. &
\tiny Need compiler wrapper to filter out options that CMake adds that break compiler. some bugs with gcc
lib compatibility &
\tiny Good integration with CMake and make.
\\

\small Hardware Mapping  &
\tiny Supports multiple host-parallel backends, and NVIDIA, AMD, and Intel GPUs.
Kokkos developers have been pro-active in supporting new hardware architectures as they emerge &
\tiny Supports multiple host parallel backends, and NVIDIA and AMD GPUs &
\tiny Can build and run now on any major vendor CPU and GPU. Third-party libraries can be called through interoperability. Backed by Intel, Codeplay, academic institutions and labs. &
\tiny nvc++ supports CPU serial, CPU multicore, NVIDIA GPU. CPU multicore doesn't work reliably.  &
\tiny Supports multiple host-parallel backends, and NVIDIA, AMD, and Intel GPUs.
\\

\hline
\end{tabularx}
\caption{Table of metrics}
\label{tab:metrics_1}
\end{table*}

\begin{table*}[ht]
\centering

\begin{tabularx}{0.99\textwidth}{ 
| >{\raggedleft\arraybackslash\hsize=2cm}X
| >{\raggedright\arraybackslash}X
| >{\raggedright\arraybackslash}X
| >{\raggedright\arraybackslash}X
| >{\raggedright\arraybackslash}X
| >{\raggedright\arraybackslash}X
|
}
\hline
\textbf{Metric} & 
\textbf{Kokkos} & 
\textbf{alpaka} & 
\textbf{SYCL} & 
\footnotesize \textbf{std::par/nvc\tiny++} & 
\textbf{OpenMP} \\
\hline

\small Feature Availability  &
\tiny Concurrent kernel only with CUDA backend, and then requires CUDA specific features. Concurrent calls to Serial backend safe, but not efficient. 
Unsupported: sort function called from device code; common API to vendor-optimized FFT libraries; RNGs not following Gaussian or uniform distributions, such as binomial &
\tiny Similar to CUDA &
\tiny Support for reductions, kernel chaining, callbacks. Concurrent kernels not supported in practice. No native support of host-parallel &
\tiny No low level controls of hardware or kernel launch parameters &
\tiny Scan and memset ops not yet supported on GPUs by most compilers. Some CUDA atomic ops also not yet supported.
\\

\small Long Term Sustainability &
\tiny Good prospects now, but what happens if CUDA/SYCL get integrated into C++ standards? Is there still a need for Kokkos? &
\tiny Possibly problematic - very small user and developer community &
\tiny Don’t foresee SYCL disappearing; essentially an OpenCL successor.
Long term support for technologies by hardware vendors. &
\tiny Part C++ standard &
\tiny Very well supported by industry
\\

\small Compilation Time &
\tiny Each backend/configuration needs its own binary. 
Does not take long to compile Kokkos libs, but kernel compilation slower than CUDA. &
\tiny Similar to native CUDA/HIP &
\tiny Similar to CUDA for small kernels. Can be prohibitively slow for very large ones &
\tiny Much slower than gcc &
\tiny Similar to CUDA
\\

\small Ease of Debugging &
\tiny Compiler error messages are difficult to read due to long template names &
\tiny Compiler error messages are very long, and not easy to decipher &
\tiny Intel VTune and Advisor are useful debugging and profiling tools. For third-party library calls, depends on the vendor (Nvidia is great, AMD is not). &
\tiny Runtime errors mostly have same name, making identification difficult. Can do some debugging with cuda-gdb. &
\tiny Clang provides runtime environment variables that help with debugging.
\\

\small Interoperability &
\tiny Kokkos and native (eg CUDA/SYCL/HIP) code can be mixed within one application. Can have issues with external libraries like TBB &
\tiny Can be mixed with CUDA/HIP and other libraries in the same application &
\tiny Good support for TBB and OpenMP &
\tiny Good support for external libraries &
\tiny Good support with external libraries and other portability layers
\\
\hline
\end{tabularx}
\caption{Table of metrics (continued)}
\label{tab:metrics_2}

\end{table*}

\subsection{Kokkos}

Kokkos provides a programming model that has a higher level of abstraction than that of CUDA. We find the learning curve to be largely similar to learning CUDA, though understanding the proper mapping of some advanced CUDA constructs, such as shuffles among threads, to Kokkos is challenging. Kokkos' runtime library requires explicit initialization and finalization calls, which can be intrusive to code organization. When using the CUDA backend, all source files that includes a Kokkos header must be compiled by the CUDA aware compiler (nvcc or clang). If the application uses shared libraries, Kokkos itself must be built with this feature enabled, and all Kokkos code in the application must be packaged into a single shared library.

We find that for relatively simple and long-running kernels, reaching the performance of the native CUDA versions is easy to achieve with Kokkos, but with complex and short kernels, the optimization requires significant effort and even then does not always reach the performance of native CUDA. There can be significant overheads from intializing Kokkos Views. We also find that concurrent use of Kokkos' API depends on the backend. For example, concurrency outside of Kokkos works with CUDA and HIP backends, whereas pthreads backend explicitly disallows more than 1 calling thread, and the serial backend serializes all API calls. 

Plentiful documentation and tutorials for Kokkos exist, as well as a very active Slack channel. The Kokkos developers' support for any kind of questions is excellent.

\subsection{SYCL}
Like Kokkos, SYCL can currently target all available backends from the same source, though recompilation is required using different versions of the compiler (openSYCL for AMD, llvm/dpc++ for NVIDIA, oneAPI for Intel). It provides comparable performance to CUDA or HIP. Programmatically, we find it to be more verbose than CUDA, but similar to Kokkos for memory management when using buffers, with similar learning curves. Its ability to automatically migrate data to devices based on data dependencies of kernel/buffer associations when using USM can simplify code design. Its support of data object types that can be offloaded is more stringent than that of CUDA. There is currently no support for concurrent kernel execution, and callback functionality is likely to be deprecated. SYCL has good interoperability with external concurrency layers such as TBB, OpenMP and MPI, though it will not natively target host parallel devices. While the SYCL standard supports concurrent kernel execution on devices, it appears that this is not currently implemented by compilers. It has mostly seamless integration with build systems like CMake. 

SYCL is strongly supported by Intel, who is pushing various features towards integration into the C++ standard. Intel offers many training classes for SYCL, and provides tools for migration from CUDA. There is a wide variety of online documentation available.

\subsection{OpenMP}
We find that offloading a C\texttt{++} kernel to GPUs via OpenMP is easy to implement and does not require major changes to the code. The architecture agnostic compiler directives can, in principle, offload to multiple GPUs and FPGAs, and compilers are under active development by NVIDIA (nvc++), AMD (rocm-clang, aomp), Intel (icpx), LLVM Clang and GCC. However, extracting performance requires fine tuning the data transfers as the \verb|map| clause implicitly transfers some variables. We also find that performance currently varies from compiler to compiler. Many specialized operations (e.g. \verb|atomic|) are currently less performant than CUDA and some operations like scan and \verb|memset| are not supported on GPUs. Manually parallelizing nested loops almost always outperforms \verb|collapse| as the latter will use more registers which degrades performance. Lastly, it is also important to tune number of threads per team and use suitable compiler flags as they sometimes drastically improve the performance. Figure \ref{fig:wct} shows the comparison of Wire-Cell Toolkit's OpenMP port with Kokkos and SYCL and Figure \ref{fig:fcs_omp} compares the performance of FastCaloSim's OpenMP port with CUDA for various GPUs.

Examples and documentation for OpenMP/offload are somewhat sparse, especially for more advanced features. Debugging and using performance tuning tools is challenging due to the extra OpenMP code infrastructure, and the architecture-specific plugins that it loads at initialization, which cause issues with NVIDIA's compute\_sanitizer and AMD's omnitrace profiler.

\subsection{alpaka}
In general, alpaka offers a lower-level API than other portability layers (e.g., Kokkos). As a result, the application code written in alpaka tends to be rather verbose. Hence, in the case of applications with a large code-base, it can be desirable to implement a shallow layer of template functions on top of the alpaka API in order to hide this complexity from the application code. 

The alpaka style of GPU code development is somewhat complex. The API is not always intuitive, which makes the learning process challenging. However, once developers are familiar with the API, porting the existing kernel code (e.g., written in CUDA) to alpaka is quite simple, as the kernel body remains practically unchanged.

As it can be seen in Figure \ref{fig:fcs_alpaka}, the FastCaloSim application performance with the CUDA back-end of alpaka is comparable with the native CUDA implementation. The slight performance degradation of the memory transfers can be explained by the fact that alpaka numbers include one extra copy of memory buffers on the host.

Alpaka has a very small developer and user community. Documentation and example code is sparse, but the developers are very active and eager to help.

\subsection{std::par}
NVIDIA's implementation of the ISO standard for parallelism (nvc++) has not been intended to be a direct replacement for CUDA. It lacks the ability to access many low level features of GPU programming, such as synchronization, explicit launch parameter control, asynchronous operations, or explicit memory transfers. Yet because it utilizes standard C++, it provides a very low entry bar for developers, offering a simplified path for porting from serial CPU code to GPUs, without the need to explicitly manage memory transfers.

While, in theory, nvc++ is link compatible with gcc libraries, there are certain limitations. Any memory that will be offloaded to the device must be allocated in code compiled by nvc++, and there are issues with some STL containers. The linking of the final executable must also be performed by nvc++. The compiler is new enough that critical bugs are still being found, though are often rapidly patched. Furthermore, it is not fully recognized by many build systems, requiring significant user intervention for more complex installations, especially when some packages require g++ and others nvc++. The compilation time with nvc++ tends to be significantly longer than with the other portability layers.

nvc++ uses Thrust to implement parallelism. When there are well matched Thrust algorithms for the parallelized regions, performance can match, or sometimes exceed equivalent CUDA code. But this performance drops off when there is not a good matching to Thrust. Unfortunately, there is no way to see the intermediate Thrust code that the compiler generates. The use of page faults and Unified Shared Memory to exchange data between device and host can also degrade performance as compared with an explicit memory transfer.

\section{Conclusions}
Portable parallel programming layers have seen tremendous development over the past several years. The range of enabled backends has vastly increased, with most now being supported by the majority of APIs. The compilers are being actively developed, with missing features being addressed, and improved performance profiles. There are strengths and weaknesses to each product, and careful choices should be made when choosing an API so that it works well with the original code and software environment, the build system, and the desired hardware. All APIs have a learning curve, though std::par is closest to normal C++, providing an easy transition path to GPU programming for little effort.  In many cases, performance of all portability layers can approach, and sometimes even exceed, that of "native" compilers, though performance tuning can come at the expense of portability.

The developers of Kokkos, SYCL, and nvc++ are aware of the benefits that standards bring, and have all made proposals to the C++ standards committees for the inclusion of various programming paradigms into the C++23 and C++26 standards. We fervently hope that these proposals succeed, and that compiler developers and hardware manufacturers collaborate to integrate them in their next generation of products, as a standards-based approach will have the best chance of long-term survival.

\section{Acknowledgments}

This work was supported by the DOE HEP Center for Computational Excellence at Brookhaven National Laboratory, and Lawrence Berkeley National Laboratory under B\&R KA2401045.
 This work was also done as part of the offline software research and development programme of the ATLAS and CMS Collaborations, and we thank the collaborations for its support and cooperation.
 This research used resources of the National Energy Research Scientific Computing Center (NERSC), a U.S. Department of Energy Office of Science User Facility located at Lawrence Berkeley National Laboratory, operated under Contract No. DE-AC02-05CH11231.
 We gratefully acknowledge the use of the computing resources provided by Brookhaven National Laboratory and by the Joint Laboratory for System Evaluation (JLSE) at Argonne National Laboratory.

\section{References}
\bibliography{refs}
\bibliographystyle{unsrt}

\end{document}